\documentclass{llncs}
\usepackage[ruled]{algorithm}
\usepackage{algpseudocode}
\usepackage{xspace}
\usepackage{xcolor}
\usepackage{amsfonts}
\usepackage{amsmath}
\usepackage{graphicx}
\usepackage{hyperref}

\newcommand{\F}{Fr\'echet\xspace}

\title{Computing Optimal Graph Mappings}
\author{Maike Buchin\inst{1}\orcidID{0000-0002-3446-4343} \and
Lea Thiel\inst{1}\orcidID{0009-0006-3510-8232}}

\institute{Department of Computer Science, Ruhr University Bochum, Germany \email{\{maike.buchin,lea.thiel\}@rub.de}}

\authorrunning{M. Buchin  and L. Thiel} 

\newtheorem{beo}{Observation}
\begin{document}
\maketitle
\begin{abstract}
The weak graph distance is a distance measure for comparing two embedded graphs, which takes into account both the topology and the geometry of the graphs. The algorithm for computing this measure produces a mapping from one graph to the other that minimises the maximum distance; however, it maps all other distances only well enough to ensure that the maximum distance is not exceeded. In practical applications, it may be of interest to map the parts of the graph beyond the bottleneck as accurately as possible. We propose two different criteria for optimal mappings and also provide algorithms to compute them.
\keywords{weak graph distance, optimal mappings, \F distance}
\end{abstract}

\section{Introduction}
There are many applications that can be modeled as embedded graphs, for example, representations of road maps, protein alignment, and river networks. 
In these applications one may need to compare two such embedded graphs with one another, for example to compare a road network graph reconstructed from GPS data with an existing graph of the network. 
As different applications require different interpretations of similarity, many different measures of graph similarity have already been proposed. These include a geometric graph edit distance~\cite{cheongetal}, traversal distance~\cite{TraversalDistance}, contour tree distance~\cite{contourTrees}, graph sampling distance~\cite{graphsampling} and local persistent homology distance~\cite{localpersistanthomology}, which are also surveyed in~\cite{ContourTreeNeu}. 
We are interested in a measure that maps one graph onto the other, and in doing so considers both the geometric and topologic error of the mapping. 
For this we use the weak graph distance introduced by Akitaya et al.~\cite{akitayaetal21}. 

 The weak graph distance is based on the \F distance \cite{altgodau}, which is a well-known distance measure for curves. It provides us with a measure of what the maximum deviation would be if we were to optimally map one graph onto the other. 
 Although the computation of the weak graph distance is generally NP-complete, there are some graph classes for which the computation can be done in polynomial time. This is the case, for example, for planar, spike-free graphs, for which Akitaya et al. \cite{akitayaetal21} have introduced an algorithm. 
 In some applications, however, it would also be of interest to know exactly how we map this graph as accurately as possible onto the other; in other words, not just to compute the maximum distance and a mapping that does not exceed this maximum distance, but a mapping where the distances are as small as possible everywhere. Let us consider the application of a road network reconstructed from GPS data: 
 we would like to assign all the data to the routes as accurately as possible, rather than basing this solely on the largest measurement error in our data. 
\begin{figure}[t]
 	\begin{center}
 		\includegraphics[scale=0.7]{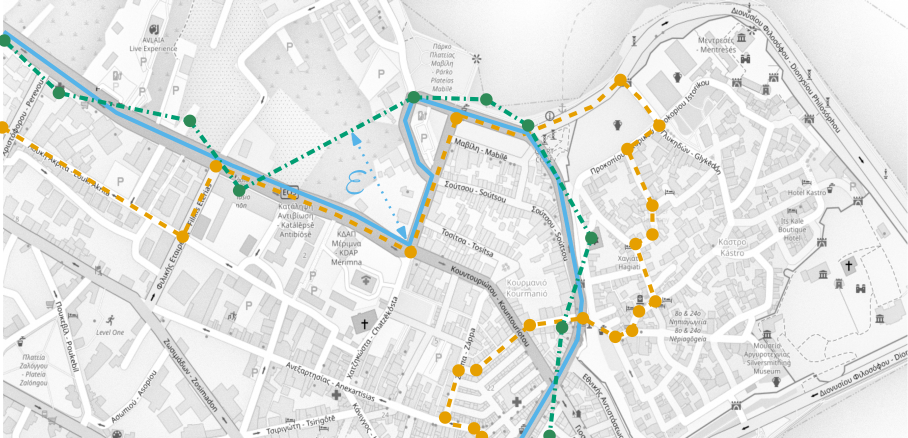}
 	\end{center}
 	\caption{A vehicle’s motion data (dash-dotted) and possible mappings (dashed, normal) onto the underlying network from OpenStreetMap that satisfy the weak graph distance.}
 	\label{Einleitungsbeispiel}
 \end{figure} 

 In Figure~\ref{Einleitungsbeispiel}, for example, we see recorded movement data for a vehicle (dash-dotted) in a city, which, in one area, does not follow a road. If we now determine possible mappings of the movement data onto the road network that satisfy the weak graph distance $\varepsilon$, both the normal route – which deviates from the data only at the bottleneck – and the dashed route – which takes many detours – are viable options. We are therefore interested in calculating a mapping that both satisfies the weak graph distance and, away from the bottleneck, represents the data as accurately as possible.
 Such an approach has already been considered for the \F distance on curves \cite{lexicographicfrechet}, \cite{LocallyCorrectFrechet}.
 
However, which mappings are regarded as optimal may depend on the specific application. Therefore we introduce two different definitions of optimal mappings for the weak graph distance: ‘edge-based optimal graph mappings’, in which the maximum distance for each edge to its image is minimised, and a lexicographic graph distance, which extends a previous definition proposed by Buchin and Kilgus~\cite{lexicographic}. 
For both approaches, we present algorithms which, for planar, spike-free graphs, compute the corresponding optimal mappings in polynomial time. Both algorithms follow the same idea, but the analysis for the lexicographic graph distance is more interesting. 
Due to space constraints, 
details for the edge-based optimal graph mappings are moved to the appendix.

\section{Preliminaries}
In this paper, we consider
connected planar graphs embedded in $\mathbb{R}^2$. Let $G_1=(V_1,E_1)$ and $G_2=(V_2,E_2)$ be two such graphs with vertices embedded as points in $\mathbb{R}^2$ that are connected by crossing free straight-line edges.

Several graph distance measures (\cite{akitayaetal21},~\cite{contourTrees},~\cite{TraversalDistance}) are based on a well-known measure for determining distances between curves: the \emph{Fréchet distance}~\cite{altgodau}.  

\begin{definition}
    Let $P\colon [a,a']\rightarrow \mathbb{R}^2$ and $Q\colon[b,b']\rightarrow \mathbb{R}^2$ be curves. The \textbf{(strong) Fréchet distance}  $\delta_F(P,Q)$ is defined as
	\begin{align*}
		\delta_F(P,Q):= \inf_{\substack{\alpha:[0,1]\rightarrow[a,a']\\\beta:[0,1]\rightarrow[b,b']}}\ {\max_{t\in [0,1]}{\ ||P(\alpha(t))-Q(\beta(t))}||}
	\end{align*}
	where $\alpha$ and $\beta$ range over all continuous and increasing functions with $\alpha(0)=a$, $\alpha(1)=a'$, $\beta(0)=b$ and $\beta(1)=b'$.
	We call $\alpha$ and $\beta$ (continuous) traversals of $P$ and $Q$.
    If we do not require monotonicity for $\alpha$ and $\beta$, we obtain the \textbf{weak Fréchet distance} $\delta_{wF}(P,Q)$. 
\end{definition}

The \emph{strong} resp.\ \emph{weak graph distance}, introduced by Akitaya et al.~\cite{akitayaetal21}, is also based  on the (strong resp.\ weak) Fréchet distance. The idea  
is to map one of the graphs onto a subgraph of the other that is as similar as possible.   
For this, a mapping $s\colon G_1\rightarrow G_2$ is called a \emph{graph mapping}	if
	it maps each vertex $v\in V_1$ to a point $s(v)$ on an edge of $G_2$, and
	it maps each edge $\{u,v\}\in E_1$ to an arbitrary simple path from $s(u)$ to $s(v)$ in the embedding of $G_2$.
    
\begin{definition}
	We define the \textbf{(directed) [weak] graph distance} $\delta_{[w]G}$ as
	\begin{align*}
		\delta_{[w]G} (G_1,G_2) = \inf_{s:G_1\rightarrow G_2} \max_{e\in E_1} \delta_{[w]F}(e,s(e)),
	\end{align*}
	where $s$ ranges over all graph mappings from $G_1$ to $G_2$, $\delta_{[w]F}$ denotes the [weak] Fr\'echet distance, and $e$ and its image $s(e)$ are interpreted as curves in the plane. 
\end{definition}

Next we briefly describe the decision algorithm for the weak graph distance from~\cite{akitayaetal21}, on which our algorithms for computing an optimal mapping are based. To do this, we first need to introduce some notation: 

\begin{definition}
	An \textbf{$\varepsilon$-placement $C_v$ of a vertex} $v$ is a maximally connected part of $G_2$ restricted to the $\varepsilon$-ball $B_\varepsilon(v)$ around $v$. An \textbf{$\varepsilon$-placement of an edge} $e=\{u,v\}\in E_1$ is a path $P$ in $G_2$ connecting placements of $u$ and $v$ such that $\delta_{wF}(e,P)\leq \varepsilon$. In this case we say that $C_u$ and $C_v$ are reachable from each other. An \textbf{$\varepsilon$-placement of the graph $G_1$} is a graph mapping $s:G_1\rightarrow G_2$ such that $s$ maps each edge $e$ of $G_1$ to an $\varepsilon$-placement.
    
	An $\varepsilon$-placement $C_v$ of a vertex $v$ is \textbf{valid} if for every neighbor $u$ of $v$ there exists an $\varepsilon$-placement $C_u$ of $u$ such that $C_v$ and $C_u$ are connected by an $\varepsilon$-placement  of the edge $\{u,v\}$. Otherwise, $C_v$ is invalid.
\end{definition}

We require the following property for graphs, which was introduced by \cite{akitayaetal21}: 
\begin{definition}
     Let $G$ be a plane graph, $C$ a simple cycle in $G$ and $\varepsilon>0$. We say that $C$ is \textbf{spike-free} (regarding $\varepsilon$), if $C$ is a cycle where both sides of a minimum size bounding box have lengths greater than $2\varepsilon$  and if  $T_\varepsilon(uw)\cap B_\varepsilon(v)=T_\varepsilon(vw)\cap B_\varepsilon(u)=\emptyset$  for all triples of vertices $u$, $v$, $w$ on $C$ such that $u$ and $v$ are adjacent to $w$ and $u$ and $v$ are not adjacent.  We call $G$ spike-free (regarding $\varepsilon$) if all cycles of $G$ are spike-free.
\end{definition}
Note that if a graph is spike-free for a value $\varepsilon$, it is also spike-free for any value $\tilde{\varepsilon}<\varepsilon$. 
The following four steps describe the decision algorithm for a fixed value $\varepsilon$ and their running time if $G_1$ is spike-free. Further details can be found in~\cite{akitayaetal21} and in the appendix.

  \begin{algorithm}[H]
	\caption{General Decision Algorithm.}
	\begin{algorithmic}[1]
		\State Compute vertex placements \Comment{$\mathcal{O}(n_1m_2) \hspace{2.5pt}$}
		\State Compute reachability information for vertex placements \Comment{$\mathcal{O}(m_1m_2)$}
		\State Prune invalid placements \Comment{$\mathcal{O}(m_1m_2)$}
		\State Decide if there exists a placement for the whole graph $G_1$ \Comment{$\mathcal{O}(1) \hspace{20pt}$}
	\end{algorithmic}
\end{algorithm}

In total this algorithm takes $\mathcal{O}(m_1m_2)$ time if $G_1$ is spike-free. Since we are considering planar graphs, we have $\mathcal{O}(m_1m_2)=\mathcal{O}(n_1n_2)$.

To compute the weak graph distance, we proceed in a similar way to the (weak) Fréchet distance: we use parametric search via critical values and the decision algorithm. According to~\cite{akitayaetal21}, the critical values are as follows:
\begin{itemize}
    \item Two vertex placements merge: the vertex in $G_2$ at which the placements are connected is of distance $\varepsilon$ from a vertex in $G_1$.
	\item A new vertex placement is created: an edge in $G_2$ is at a distance of $\varepsilon$ from a vertex in~$G_1$.
    \item An edge placement may emerge: the weak Fréchet distance from a path to an edge is $\varepsilon$, i.e., a vertex of $G_2$ is $\varepsilon$ away from an edge of $G_1$.
\end{itemize}

There are $\mathcal{O}(n_1m_2)$ values of the first two types and $\mathcal{O}(m_1n_2)$ of the third type, so that the total running time on planar graphs is $\mathcal{O}(n_1n_2\log(n_1+n_2))$.

\section{Optimal Weak Graph Distance}
The exact criteria to determine which mapping for the weak graph distance is ‘optimal’ may depend on the underlying setting.
   \begin{figure}[t]
 	\begin{center}
 		\includegraphics[scale=1]{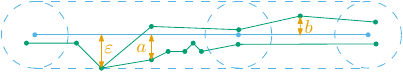}
 	\end{center}
 	\caption{The two edges can be mapped onto the graph in two ways, either to the path above or below the right edge. It holds that $\varepsilon > a > b$.}
 	\label{Verschiedene_Wahlen}
 \end{figure} 

 In Figure~\ref{Verschiedene_Wahlen}, for example, we see two simple graphs whose weak graph distance is $\varepsilon$. However, there are two ways to map the graph $G_1$ onto $G_2$ with minimum distance. Since the weak graph distance corresponds to the smallest possible maximum weak \F distance between an edge and its image, the first natural approach is to define an optimal mapping that minimises all these distances.  In the example, the weak graph distance $\varepsilon$ is attained for the first edge of $G_1$. So, regardless of which mapping we choose, the distance of this edge to its image will always be $\varepsilon$. If we now aim to minimise the weak \F distance to its image for the second edge of $G_1$, we choose the lower of the two possible paths for the mapping in order to avoid the bottleneck of width $b$. 
 However, if for each edge we want to not only minimise its maximum weak \F distance but also ensure that each edge is mapped as well as possible overall, 
 we choose the upper path as the optimal one in this example. 
 The distance at all other points, apart from the outlier which realises the weak \F distance of the first edge to the possible images, is smaller for the upper path. So, when considering the remaining options for a mapping as a whole, one would prefer to avoid distance $a$ rather than distance $b$, since $a>b$. 
 
 In the following we will therefore present two natural definitions of an optimal mapping for the weak graph distance, which realise the two different ideas of optimality from the example, and also give algorithms for computing the corresponding optimal mappings in polynomial time. 
 Both algorithms compute a distance vector for an optimal mapping rather than just a single distance value. For the lexicographic graph distance, we avoid recomputing the weak graph distance from scratch at each step by sweeping the critical values in decreasing order and reusing the decision procedure of \cite{akitayaetal21}; this keeps the overhead for computing the weak graph distance itself to a factor of $n^2/\log(n)$ in the worst case, and to $n/\log(n)$ whenever the distance vector has linear size. For the edge-based optimal mapping, the overhead is a factor of $n$. The corresponding mapping can then be reconstructed from the critical values of the minimum distance vector. 
 
In the following, we only consider graphs for which the weak graph distance can be computed in polynomial time using the algorithm in~\cite{akitayaetal21}. This means that both $G_1$ and $G_2$ are planar, and $G_1$ is also spike-free. Furthermore, we assume that all critical distances are unique.

\subsection{Lexicographic Graph Distance}
We now want to formalise the idea of not only minimising the maximum distance for each edge, but also representing each edge as accurately as possible.
Buchin and Kilgus~\cite{lexicographic} proposed a definition for this type of optimal weak graph distance, along with an algorithm for computing this optimal mapping. 
However, their definition does not quite correspond to the intuition of distance that they describe, see the Appendix for further details. Below, we provide a definition that captures the desired intuition behind the optimal mapping. The optimal mapping should satisfy the following property: whenever a part of the graph $G_1$ can be mapped locally better than with its current distance $\delta$, this causes another distance to increase to a value greater than $\delta$. 

Let us illustrate, using the example in Figure \ref{Verschiedene_Wahlen}, how we can optimally map a single edge onto a graph $G_2$ and which choices we need to consider. If we compute the weak graph distance, there are already three decisions that need to be made:
The weak graph distance in the example corresponds to $\varepsilon$, which means that the path onto which our edge is mapped must pass through the vertex to which this critical value belongs. Furthermore, the vertices of our edge from $G_1$ must be mapped onto one of the remaining vertex placements. In this case, any position on the respective placements is possible for both vertices.
To decide which of the remaining options is optimal for an edge placement, let us make use of the following observation from \cite{akitayaetal21}: the weak \F distance between a single edge $e$ and a path $s(e)$ corresponds exactly to the maximum of the Hausdorff distance between $e$ and $s(e)$ and the distances between the respective endpoints of $e$ and $s(e)$.  We therefore want to select, from the subpaths available to us, those whose Hausdorff distance to our edge is the smallest, so as to achieve optimal mapping even outside the bottleneck. In our example, this means that we want to avoid the vertex with distance $a$ and choose the other subpaths respectively in order to place our edge.  The relevant values for determining the optimal placement of the edge are therefore both the Euclidean distances of the vertices in $G_1$ from all vertices and edges of $G_2$ for the possible vertex placements, and the shortest distance from each vertex on $G_2$ to an edge of $G_1$, in order to determine the optimal path. We therefore define the critical distances of a mapping as follows:

 \begin{definition}
 	Let $s\colon G_1\rightarrow G_2$ be a graph mapping. Let $e$ be an edge on $G_1$ and $s(e)$ be the path on $G_2$ to which $e$ is mapped. Then we call the shortest distances from every vertex on $s(e)$ to $e$ the \textbf{critical distances of $e$ with respect to~$s$}.
 	We denote the distances $||u-s(u)||$ for $u\in V(G_1)$ and the critical distances of all edges of $G_1$ with respect to $s$ as the \textbf{critical distances of $s$}. We call the vector $\delta$ of the descending order of the critical distances of $s$ the \textbf{distance vector of $s$}.
 \end{definition}
 Note that the endpoints $s(u),s(v)$ of $s(e)$ only contribute to the critical distances of an edge $e$ if they are vertices of $G_2$. Otherwise, they are counted based on the distances from the vertices of $G_1$ to their images.
 As we want to optimise not for a single edge but for all of them together in order to achieve optimal mapping, we define optimal mapping as follows:
 
 \begin{definition}
 	Let $s\colon G_1\rightarrow G_2$ be a graph mapping with distance vector $\delta$.  Then \textbf{$s$ realises the lexicographic graph distance} between $G_1$ and $G_2$ if, for every other mapping $\tilde{s}$ with distance vector $\tilde{\delta}$, we have $\delta\leq_{lex}\tilde{\delta}$. Distance vectors of different length are compared by padding the shorter one with zeros.
 \end{definition}

 The optimal mapping corresponds to the lexicographically smallest possible vector and thus satisfies the intuition described above: if we were able to improve the mapping further by locally improving a distance, it would no longer be the lexicographically smallest vector, and consequently a higher distance value would have to be increased further. To calculate such an optimal mapping, we now make use of the following observation, the proof of which can be found in the appendix: 

\begin{lemma}
    The possible critical distances of an optimal mapping correspond to the critical values used to compute the weak graph distance. And the first entry of the distance vector corresponds to the weak graph distance.
\end{lemma}

The distance vector of a mapping $s$ therefore contains those critical distances that are attained for the mapping.
Since there are  $\mathcal{O}(n_1n_2)$ critical values for the computation of the weak graph distance, the distance vector $\delta$ of a mapping can also have at most $\mathcal{O}(n_1n_2)$ entries. As can be seen in the Appendix, the vector can reach a size of $\Theta(n_1n_2)$, but we expect it to be much smaller in practice.

\subsection{Computation of the Lexicographic Graph Distance}
The computation algorithm by Buchin and Kilgus \cite{lexicographic} for their variant of the lexicographic graph distance uses the following strategy:
at each step, the weak graph distance is computed using the algorithm described in \cite{akitayaetal21}, and the bottleneck at which the weak graph distance is attained is then removed. 

To remove the bottlenecks, they store a separate copy of $G_2$ for each edge of $G_1$, so that each vertex in $G_2$ at which the current graph distance is attained can be moved to the corresponding edge of $G_1$. 
As this takes place in a separate copy of $G_2$, it remains possible to map other edges of $G_1$ onto this part of $G_2$.  
Since the weak graph distance is computed for each of the $\mathcal{O}(n_1n_2)$ entries of the distance vector $\delta$, the approach requires a total running time of $\mathcal{O}(n_1^2n_2^2\log(n_1+n_2))$. 

Thus in their algorithm, the decision algorithm  may be executed multiple times for some values of $\varepsilon$ due to the parametric search (on graphs that may have changed). However, it is not straightforward to reuse these previous calculations for a given $\varepsilon$, as the shifting of previous bottlenecks may result in more valid vertex and edge placements than before. 
Overall, it is therefore more efficient to calculate the weak graph distance only once and then test each critical value decreasing in size. In addition, we can then make use of the following observation:
\begin{beo}
	If the decision algorithm for the weak graph distance returns the answer $\delta_{wG}(G_1,G_2)\leq \varepsilon$, then all invalid placements removed in this pruning step are invalid for any smaller $\varepsilon$.
	\label{Beo:bleibt_gelöscht}
\end{beo}
However, unlike in the decision algorithm from \cite{akitayaetal21} for the weak graph distance, we do not want to recompute all possible vertex placements at every step. 
So, rather than computing the vertex placements anew at every step, for each vertex $v\in V_1$, we maintain a list $L_v$ of its vertex placements that are still valid. From this list, the vertex placements valid for the new critical value can then be computed the next time our modified decision algorithm is run.  
 For convenience, when we apply this modified decision algorithm to compute the weak graph distance with restricted vertex placements, we simply refer to this as deciding the weak graph distance as well. 
  With at most $n_1n_2$ critical values, we obtain a runtime of $\mathcal{O}(n_1^2n_2^2)$.
  The following algorithm calculates the lexicographic graph distance using the modified decision algorithm in line 10:

\begin{algorithm}[H]
  	\caption{Lexicographic Graph Distance}
    \hspace*{\algorithmicindent} \textbf{Input:} Graphs $G_1$ and $G_2$
  	\begin{algorithmic}[1]
  			\State Compute critical values $\varepsilon_1, \ldots, \varepsilon_\ell$ for the weak graph distance, sorted in descending order.
            \State $\varepsilon_{\ell+1}=0$
            \State Compute $\delta_{wG}(G_1,G_2)=\varepsilon_i$. 
            \State For each edge in $G_1$, store the intersection of the $\varepsilon_i$-tube with $G_2$. Remove all invalid connected components in these.
            \State For each vertex $v\in G_1$, store the vertex placements $L_v$ that are still valid and the connected components to which they belong.
            \State $\delta_1\leftarrow\varepsilon_i$
            \State \textbf{Procedure:} Remove  bottleneck $\varepsilon_i$.
  			\State $j\leftarrow2$
  			\While{$i\leq \ell+1$}
            \State Decide if the current weak graph distance $\delta_{wG}$ is greater than $\varepsilon_i$ using the  \hspace*{\algorithmicindent} stored list of valid vertex placements and the distance parameter $p$.
  			\If {  $\delta_{wG}> \varepsilon_i$}
  			\State \textbf{Procedure:} Remove  bottleneck $\varepsilon_{i-1}$.
  			\State $\delta_j\leftarrow\varepsilon_{i-1}$
  			\State $j++$
  			\EndIf
  			\State $i++$
  			\EndWhile
	\end{algorithmic}
     \hspace*{\algorithmicindent} \textbf{Output:} distance vector $\delta=(\delta_1,...,\delta_{j-1})$.
     \label{Algo:lexicographic}
\end{algorithm}
   
In order to remove the bottlenecks in each step, we need to distinguish between the different types of critical values: 
\begin{itemize}
   	\item{Case 1:} Distance of a vertex $w$ of $G_2$ to a vertex $v$ of $G_1$.
   	\item{Case 2:} Distance of an edge $(w_1,w_2)$ of $G_2$ to a vertex $v$ of $G_1$.
   	\item{Case 3:} Distance of a vertex $w$ of $G_2$ to an edge $(v_1,v_2)$ of $G_1$.
\end{itemize}

 To remove these bottlenecks, we use the basic idea proposed by Buchin and Kilgus~\cite{lexicographic}, if the corresponding critical value for the bottleneck is the distance between a vertex $v$ in $G_1$ and $G_2$, i.e. the concrete placement for a vertex from $G_1$ is determined:  If the corresponding point on $G_2$ is a vertex, it is moved to $v$ and stored as the only placement of $v$. If the corresponding point is on an edge of $G_2$, we create an auxiliary vertex, as shown in Figure~\ref{Beispiel_Knoten_snappen}, and move it to $v$. 
Note that whilst these steps do affect the planarity and the number of edges and vertices in the graphs, they do not impact the runtime of the algorithm, as we continue to work with the stored placements. However, in case $3$, rather than actually moving the vertices of $G_2$ onto the edges of $G_1$, we do so only symbolically. To do this, for each copy of the vertices of $G_2$, we use a distance parameter $p$, which initially stores the actual distance to the corresponding edge of $G_1$ and is set to $0$ as soon as it is to be moved symbolically to the edge of $G_1$. So, when using the decision algorithm for the weak graph distance, if we wish to determine which vertices of $G_2$ lie within the current $\varepsilon$-tube around an edge of $G_1$, these are all vertices with $p \leq \varepsilon$. The following procedure summarises the approach:

   \begin{figure}
   	\begin{center}
   		\includegraphics[]{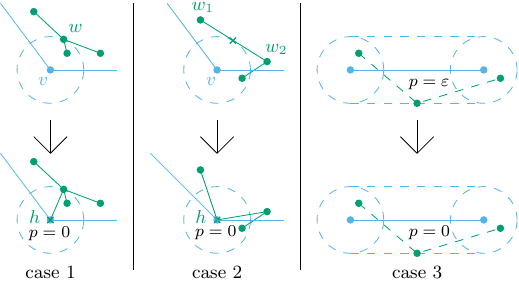}
   	\end{center}
   	\caption{Examples of how to remove the three different types of bottlenecks.}
   	\label{Beispiel_Knoten_snappen}
   \end{figure}  
  \begin{algorithm}[H]
 	\caption{Procedure: Remove bottleneck $\varepsilon$ }
 	\begin{algorithmic}[1]
	 		\If{the distance $\varepsilon$ is attained between vertex $v\in V_1$ and vertex $w\in V_2$}
	 		\State Create an auxiliary vertex $h\in V_2$ at the position of $v$ with $p=0$ and an    
            \hspace*{\algorithmicindent}
            edge $(h,w)$ in all copies of the edges incident to $v$.
	 		\State Store the vertex $h$ as the only vertex placement for $v$.
	 		\ElsIf{$v$ is mapped to an edge $(w_1,w_2)\in E_2$}
	 		\State Create an auxiliary vertex $h$ at the position of $v$ with $p=0$ on each edge  
            \hspace*{\algorithmicindent}incident to $v$.
	 		\State Create edges $(h, w_1)$ and $(h, w_2)$ in all copies of the edges incident to $v$.
	 		\State Delete the edge $(w_1, w_2)$ from all copies of edges incident to $v$. 
	        \State Store the vertex $h$ as the only vertex placement for $v$. 
            \Else 
	 		\State Set the parameter $p$ of $v$ to $0$.
	 		\EndIf
	 		\State Delete all other connected components in all modified copies of $G_2$.
            \State Run the pruning step, update lists $L$ of valid placements.
	 	\end{algorithmic}
\end{algorithm}

Once this procedure has removed a bottleneck, the graphs are adjusted so that the remaining critical values of the optimal mapping remain  unchanged:
\begin{lemma}\label{Lemma-Korrektheit-Eliminierung}
	Let $s\colon G_1\rightarrow G_2$ with distance vector $\delta=(\delta_1,...,\delta_l)$ be a mapping that realises the lexicographic graph distance between $G_1$ and $G_2$. Then, after the procedure “Remove bottleneck $\delta_1$”, the lexicographic distance vector of the graph copies is $\tilde{\delta}=(\delta_2,\ldots,\delta_l)$.
\end{lemma}
\begin{proof}
	The current weak graph distance $\varepsilon$ corresponds to the first entry $\delta_1$ of the vector $\delta$ for the current graphs $G_1$ and $G_2$.
    We must show that the point at which the graph distance is measured is adjusted so that the distance at that point becomes $0$ and that the  remaining distance vector is not changed.
    We first consider case $1$ (see left part of Figure ~\ref{Beispiel_Knoten_snappen}), in which a vertex $v\in V_1$ is placed on a vertex $w\in V_2$. We create an auxiliary vertex $h$ at the position of $v$ and also create an edge connecting the auxiliary vertex to the vertex $w$ onto which $v$ is placed. We can adjust a possible mapping of the computed weak graph distance as follows, thereby reducing the distance: Instead of mapping the vertex $v$ to $w$, we map $v$ to $h$. The distance between $v$ and $h$ is zero. Furthermore, the weak graph distance of each edge incident to $v$ to its respective image is reduced: since the critical values must be unique, the distance from $w$ to each edge incident to $v$ is smaller than the computed weak graph distance. We denote the point at which this critical value is attained on an edge $e$ as $b(e)$. We map the part of each edge $e$ between $v$ and $b(e)$ to the new edge $(h,w)$, and the remaining part of each edge to its original path. This removes the bottleneck, and since we do not add the auxiliary vertex $h$ as a possible placement for other vertices, the remaining distance vector is unchanged.
    
	For the second case (see middle part of Figure~\ref{Beispiel_Knoten_snappen}), we consider a vertex $v\in G_1$ which is placed on an edge $(w_1,w_2)$. We create a vertex $h$ at the position of $v$, whose distance from $v$ is therefore $0$. The original edge is deleted along with its distance $\varepsilon_1$, and, as in case $1$, the new edges no longer contribute to the distance.
    
    In the third case (see right part of Figure~\ref{Beispiel_Knoten_snappen}), only the parameter $p$ is set to~$0$, and the remaining distance vector is not changed. 

    The optimal mapping for the lexicographic graph distance between $\tilde{G_1}$ and $\tilde{G_2}$ definitely uses the point that has just been changed. Since all distance values are unique, an alternative optimal mapping here would contradict the optimality of the original mapping.
	  Since we have already decided on a specific mapping for a vertex, the corresponding edge of $G_1$ must also be mapped entirely onto the corresponding connected component of our copy of $G_2$. All other connected components are therefore no longer relevant to the calculation and can be removed without altering the distance vector.    
\end{proof}

 We obtain the following theorem, which gives us an initial bound on the runtime, which will later be improved upon. The proof is in the Appendix. 
 \begin{theorem}
	 Algorithm \ref{Algo:lexicographic} computes the lexicographic graph distance for a spike-free graph $G_1$ and a planar graph $G_2$, and an optimal mapping, in time $\mathcal{O}(n_1^2n_2^2)$. 
\end{theorem}

In what follows, we want to take advantage of the fact that both the vertex placements and the edge placements change only slightly when the critical values are checked one after the other.

When searching for the next entry in $\delta$, according to observation \ref{Beo:bleibt_gelöscht}, we only need to continue working with those placements that remain after this pruning step. Since we are testing the critical values of $\varepsilon$ in order of size, in the first two steps of the decision algorithm for $\varepsilon$, only one of the following three changes can occur – depending on the type of the previous event:
At the critical value $\varepsilon_{i-1}$...
\begin{enumerate}
	\item two vertex placements merge: the placement that merged at the event for $\varepsilon_{i-1}$ is split into two placements. All other placements remain as they are; their connectivity components also remain unchanged for the moment. The connectivity components must be redefined for the two newly created smaller placements. This may result in some placements becoming invalid; a pruning step needs only be carried out starting from the modified components.
	\item a new vertex placement is created. This placement is therefore no longer present for $\varepsilon$; everything else remains unchanged for the moment. A pruning step must be carried out starting from this placement.
	\item the weak \F distance from a path to an edge is attained. That connected component is split; connectivity for this part must be re-checked, and a pruning step may need to be carried out on the associated vertex placements.
\end{enumerate}

Overall, we 
see that it is not necessary to recompute all vertex placements and reachability information at every step; instead, changes occur only locally. 
In the following, we analyse how much time the respective pruning steps take. We observe that in some cases, the pruning steps may require quadratic runtime:
\begin{beo}
	If $\varepsilon_{i-1}$ is an entry of $\delta$, then the worst-case runtime of $\mathcal{O}(n_1n_2)$ for the pruning step in the original algorithm from \cite{akitayaetal21} can be achieved.
\end{beo}  
Since $\varepsilon_{i-1}$ is an entry in our vector $\delta$, it corresponds to the current weak graph distance of our modified graphs. The decision algorithm for the critical value $\varepsilon$ must therefore conclude that no mapping with a distance smaller than $\delta$ is possible; thus, at least enough vertex placements must be deleted so that no vertex has a valid placement remaining. As analysed in~\cite{akitayaetal21}, this step may take $\mathcal{O}(n_1n_2)$ time. 
For all other executions of the decision algorithm, we observe:
\begin{beo}
	If $\varepsilon_{i-1}$ is not an entry in the vector $\delta$, the decision algorithm for $\varepsilon_i$ also returns the answer “yes”. According to observation $\ref{Beo:bleibt_gelöscht}$, all placements that are deleted during the decision for $\varepsilon_i$ may remain deleted. Thus, these placements are deleted exactly once in all these steps combined.
\end{beo}
We can therefore prove the following lemma:
\begin{lemma}
    	Let $\mathcal{A}$ be the set of all $\varepsilon_i$ for which $\varepsilon_{i-1}$ is not an entry of the vector~$\delta$. Then the total runtime of all executions of the decision algorithm in (Line 10 of) Algorithm~\ref{Algo:lexicographic} for values from $\mathcal{A}$ is $\mathcal{O}(n_1n_2^2+n_1^2n_2)$.
\end{lemma}
\begin{proof}
	We analyse the three steps of the decision algorithm separately:
	\begin{itemize}
		\item When computing the vertex placements, changes only occur for the first two types of critical values. In one case, a placement disappears; in the other, a placement is split into two. So we can adjust these placements in $\mathcal{O}(1)$ time.
	    \item If a placement of a vertex $v\in V_1$ has split, the connectivity must be re-checked in each associated connected component. This takes $\mathcal{O}(deg(v)n_2)$ time. For a fixed vertex $v$, a placement can only be split at vertices of $G_2$, so a total of $n_2$ times; this means that, across all critical values of $\varepsilon$, splitting placements for a fixed vertex $v$ takes $\mathcal{O}(deg(v)n_2^2)$ time. Thus, for all vertices, we require a total of $\mathcal{O}(\sum_{v\in V_1}(deg(v)n_2^2))$ time; however, as our graphs are planar, we have $\mathcal{O}(\sum_{v\in V_1}(deg(v)n_2^2))=\mathcal{O}(n_1n_2^2)$.
	    If a placement disappears, it needs to be removed from the lists of vertex placements for all associated connected components ($\mathcal{O}(n_1)$).
        In the latter case, the new connected components must be computed for the corresponding copy of $G_2$ ($\mathcal{O}(n_2)$).
	    \item As described above, each vertex placement can only be removed once by pruning. Therefore, all pruning steps together take $\mathcal{O}(n_1n_2)$ time.
	\end{itemize}
	This gives a total runtime of $\mathcal{O}(|\mathcal{A}|(1+n_1+n_2)+n_1n_2 +n_1n_2^2)=\mathcal{O}(n_1n_2^2+n_1^2n_2)$.
\end{proof}
As before, we can reconstruct the mapping from the obtained distance vector, and thus get the following result for computing the lexicographic graph distance:
\begin{theorem}
	A mapping that realises the lexicographic graph distance between a spike-free graph $G_1$ and a planar graph $G_2$ can be computed in $\mathcal{O}(|\delta|n_1n_2+n_1^2n_2+n_1n_2^2)$ time, where $\delta$ is the corresponding distance vector with a maximum size of $O(n_1 n_2)$.
\end{theorem}
We typically expect $\delta$ to be of size linear in $n_1$, i.e., each edge only adds
$\mathcal{O}(1)$ distance values. Then the runtime of the algorithm is cubic in $n= \max(n_1, n_2)$.  

\subsection{Edge-based optimal weak graph distance} 
 We formalise the idea of optimising the mapping by minimising all weak \F distances between the edges and their images:
  \begin{definition}
     Let $s\colon G_1\rightarrow G_2$ be a mapping. Let $e_1,\dots,e_m$ be an ordering of the edges of $G_1$ such that  $\delta_1\geq\dots\geq\delta_m$, where $\delta_i=\delta_{wF}(e_i,s(e_i))$. We call $(\delta_1,\dots,\delta_m)$ the \textbf{edge-based distance vector} of $s$. 	
     A mapping $s$ with edge-based distance vector $\delta$ is called \textbf{edge-based optimal} if, for every other mapping $\tilde{s}$ with distance vector $\tilde{\delta}$, we have $\delta\leq_{lex}\tilde{\delta}$.
 \end{definition}
 To compute the edge-based distance vector and thus the edge-based optimal mapping between two graphs, we can once again repeatedly apply the computation algorithm for the weak graph distance and, after each computation, adjust the graph $G_1$ in such a way that the current bottleneck is eliminated and the optimal mapping can be determined for the remaining graph. Further details and the proof of the following Theorem can be found in the Appendix.
 
 \begin{theorem}
         We can compute an edge-based optimal graph mapping with associated distance vector $\delta$ in $\mathcal{O}(n_1^2n_2\cdot \log(n_1+n_2))$ time. 
         \label{wichtig}
 \end{theorem}

\section{Conclusion}
We have presented two ways in which mappings that satisfy the weak graph distance property and, in addition, map the graph as well as possible throughout can be computed. In doing so, we increase the runtime required for any mapping that satisfies the weak graph distance property by a factor of $n$ or $n^2/\log(n)$ respectively. If our expectation — that typically the distance vector is linear in size —  is correct, the second runtime increase would even be reduced to $n/\log(n)$. 

It would be interesting to test, both experimentally and on realistic data such as road maps, what size the distance vector actually reaches. It would also be interesting to investigate whether a small distance vector can be guaranteed for certain graph classes, or whether further properties of the graphs can be utilised to reduce the runtime. 

\bibliography{bibtex.bib}
\newpage
\appendix
\section{Decision algorithm for the weak graph distance}
We give a brief explanation of the decision algorithm for the weak graph distance. 
\vspace*{-0.5cm}
\setcounter{algorithm}{0}
  \begin{algorithm}[H]
	\caption{General Decision Algorithm.}
	\begin{algorithmic}[1]
		\State Compute vertex placements \Comment{$\mathcal{O}(n_1m_2) \hspace{2.5pt}$}
		\State Compute reachability information for vertex placements \Comment{$\mathcal{O}(m_1m_2)$}
		\State Prune invalid placements \Comment{$\mathcal{O}(m_1m_2)$}
		\State Decide if there exists a placement for the whole graph $G_1$ \Comment{$\mathcal{O}(1) \hspace{20pt}$}
	\end{algorithmic}
\end{algorithm}
\vspace*{-0.5cm}
\setcounter{algorithm}{3}
\begin{enumerate}
	\item Compute vertex placements: For each vertex $v\in V$, all $\mathcal{O}(m_2)$ placements are computed. Using algorithms for computing connected components, this step takes $\mathcal{O}(n_1m_2)$ time in total.
    \item Compute reachability information for vertex placements: Iterate over all edges $e=\{u,v\}\in E_1$ to compute those pairs of vertex placements $C_u$ and $C_v$ that allow the edge $e$ to be placed. For this there must exist a path from $C_u$ to $C_v$ within the $\varepsilon$-tube around~$e$. If one determines connected components of the graph within this tube (for example, using breadth-first search starting from placements not yet found), one obtains valid edge placements for every combination of $C_u$ and $C_v$ belonging to the same connected component. For each of these components, we store a list containing the corresponding vertex placements. In total, this step takes $\mathcal{O}(m_1m_2)$ time.
	\item Prune invalid placements:  we now want to remove all vertex placements that are not valid. 
    It is possible that removing invalid placements may cause further placements to become invalid.  
    If an invalid vertex placement $C_v$ is deleted from $v$, this placement must be removed from the lists of all associated connected components for edge placements. If one of these connected components contains no further placements for $v$, then all other placements for other vertices in this connected component become invalid. This step is carried out recursively until there are no more invalid placements. In the worst case, this step takes $\mathcal{O}(m_1m_2)$ time.
	\item Decide if there exists a placement for the whole graph $G_1$: This step is generally NP-complete. However, if the graph $G_1$ is spike-free, then a placement for $G_1$ exists if and only if, after the first three steps, every vertex of $G_1$ still has at least one valid placement.
\end{enumerate}
\section{Original Definition of Lexicographic Graph Distance}
The following definition of the lexicographic graph distance was proposed by Buchin and Kilgus in \cite{lexicographic}:
\begin{definition}
	Let $s\colon G_1\rightarrow G_2$ be a mapping. We say that s is a mapping realizing the \textbf{lexicographic graph distance}, if it has the following property: Given an arbitrary subdivision $D$ of the graph $G_1$ (with finite number $l$ of edges). Let $e_1,e_2,...,e_l$ be a numbering of the edges of $D$ such that $\varepsilon_1\geq\varepsilon_2\geq\dots\geq\varepsilon_l$, where $\varepsilon_i=\delta_{wF}(e_i,s(e_i))$. If there exists another mapping $\tilde{s}$, such that $\delta_{wF}(e_i, \tilde{s}(e_i))<\delta_{wF}(e_i,s(e_i))$ for some $i\in \{1,2,\dots,l\}$, then there exists an index $j<i$ such that $\delta_{wF}(e_j,\tilde{s}(e_j))>\delta_{wF}(e_j, s(e_j))$.
    \label{old-definition}
\end{definition} 
However, for various reasons, this definition does not correspond to the intuition of distance that they describe:

\begin{itemize}
    \item 	To ensure that the graph is mapped as best as possible on each part, the property should hold for every subdivision of the graph. However, the weak graph distance changes when additional vertices are added to $G_1$; see Figure \ref{Unterteilen_geht_nicht}.
     Since additional vertices on $G_1$ must also be uniquely mapped to $G_2$, a mapping $s$ from $G_1$ to $G_2$ does not immediately allow us to construct a mapping $\tilde{s}$ from a subdivision $D$ of $G_1$ to $G_2$. In Figure~\ref{Unterteilen_geht_nicht}, the point where the new vertex is placed in $D$ is traversed a total of three times by the original mapping; however, for the mapping of $D$, it must be mapped uniquely. 
    \begin{figure}[t]
		\begin{center}
			\includegraphics[scale=1.5]{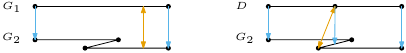}
		\end{center}
		\caption{Two graphs whose weak graph distance changes when $G_1$ is subdivided. The blue arrows show the mapping of the vertices, while the yellow ones show the weak graph distance.}
		\label{Unterteilen_geht_nicht}
	\end{figure} 
	\item If edge $e_i$ can be mapped locally better for another mapping $\tilde{s}$, then one of the $i-1$ larger distances should be increased accordingly. However, in Definition~\ref{old-definition} only edges are considered that have a larger distance in the original mapping. It should not, however, be the case that edges which were previously well-mapped produce a larger distance in the new matching; see Figure \ref{Def_nicht_hinreichend}. 
	\begin{figure}[h]
		\begin{center}
			\includegraphics[]{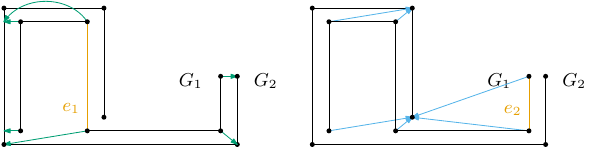}
		\end{center}
		\caption{The left-hand image shows an intuitively optimal mapping $s$, in which $e_1$ is furthest from the image. The image on the right shows a different mapping $\tilde{s}$, in which $e_1$ is mapped better; however, edge $e_2$, for example, is further from the image than $\delta_{wF}(e_1,s(e_1))$. Nevertheless, according to Definition~\ref{old-definition} the mapping on the right demonstrates that the mapping on the left is not optimal.}
		\label{Def_nicht_hinreichend}
	\end{figure} 
\end{itemize} 

\section{Proof of Lemma $1$}
\setcounter{lemma}{0}
\begin{lemma}
 The possible critical distances of an optimal mapping correspond to
the critical values used to compute the weak graph distance. And the
first entry of the distance vector corresponds to the weak graph distance   
\end{lemma}
\setcounter{lemma}{3}
\begin{proof}
    There are two types of entries in the optimal vector $\delta$: 
    \begin{enumerate}
        \item The critical distances of edges with respect to the optimal mapping: These correspond exactly to the third type of critical distances in the calculation of the weak graph distance, where an edge placement may emerge. 
        \item The distances of the vertices of $G_1$ to their placements on $G_2$: For a contradiction, we assume that the distance $\tilde{\delta}$ from a vertex $u$ to its image $s(u)$ does not correspond to a critical distance with respect to the weak graph distance. Then there exists a critical value $\varepsilon$ for which the placement containing $s(u)$ is created (Second type of critical value of the weak graph distance). We call the corresponding point in $G_2$ $s^\prime(u)$. If we map $u$ to $s^\prime(u)$, we obtain a mapping for each incident edge $e$ by connecting the path from $s^\prime(u)$ to $s(u)$ with $s(e)$. If this creates loops, they can be shortened by taking the direct path. If we retain the remainder of the mapping $s$, we obtain a valid mapping $s^\prime$. The entry for vertex $u$ was reduced because $\varepsilon$ as the critical value that created the placement must be smaller than $\tilde{\delta}$.  The path $s^\prime(u)$ to $s(u)$ lies entirely within an $\tilde{\delta}$-ball centered at $u$. Thus, although new entries of $\delta$ could be generated, they are all smaller than $\tilde{\delta}$ due to the uniqueness of the critical values. Hence both vectors agree on all entries greater than $\tilde{\delta}$, while at the position where $\delta$ has the value $\tilde{\delta}$, the vector of $s^\prime$ has a strictly smaller value. Therefore the distance vector of $s^\prime$ is lexicographically smaller, contradicting the optimality of $s$.
    \end{enumerate}
    Next we show that the first entry of the distance vector of an optimal mapping corresponds to the weak graph distance. 
    Suppose we have an optimal mapping $s$ with distance vector $\delta$ and first entry $\delta_1$. Every vertex $u$ of $G_1$ is assigned an image $s(u)$ by $s$ and all distances are less than or equal to $\delta_1$. Now, if we consider an edge $e$ and its image $s(e)$, we know that every vertex of $G_2$ in $s(e)$ can be at a distance of at most $\delta_1$ from $e$. Thus, the weak Fréchet distance from $e$ to $s(e)$ is also less than or equal to $\delta_1$. Overall, $s$ is therefore a valid mapping for the weak graph distance. It remains to be shown that $\delta_1$ is the smallest possible value for a valid mapping. 
    Suppose the weak graph distance is less than $\delta_1$, and let $\tilde{s}$ be a corresponding mapping. Every vertex $u$ is uniquely mapped, and we have $||u-\tilde{s}(u)||<\delta_1$. Furthermore, for every edge $e$, the weak Fréchet distance to $\tilde{s}(e)$ is less than $\delta_1$, and thus the distance from every vertex in $G_2$ to $\tilde{s}(e)$ is less than $\delta_1$. Thus $\tilde{s}$ is a valid mapping for the lexicographic graph distance with a smaller first entry, which leads to a contradiction.
\end{proof}

\section{Worst-Case Example of the Distance Vector for Lexicographic Graph Distance}
 \begin{figure}[H]
 	\begin{center}
 		\includegraphics[scale= 0.9]{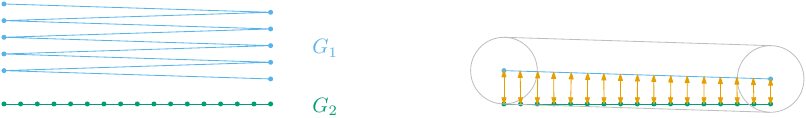}
 	\end{center}
 	\caption{Graphs $G_1$ and $G_2$, with an optimal mapping whose corresponding vector $\delta$ has $\mathcal{O}(n_1n_2)$ entries. Each edge in $G_1$ is mapped onto the entire graph $G_2$, as illustrated in the figure on the right for an edge of $G_1$. For each edge, there are $n_2$ critical values associated with the vector $\delta$.}
 	\label{Beispiel_Vektor}
 \end{figure} 
\begin{beo}
  There exist Graphs $G_1$ and $G_2$ with a mapping $s : G_1 \rightarrow G_2$ that realises the lexicographic graph distance and the corresponding distance vector $\delta$ has $\Theta(n_1n_2)$ entries (see Figure \ref{Beispiel_Vektor}).
\end{beo}

\medskip
\section{Proof of Theorem 1}
\setcounter{theorem}{0}
\begin{theorem}
	 Algorithm \ref{Algo:lexicographic} computes the lexicographic graph distance for a spike-free graph $G_1$ and a planar graph $G_2$, and thus an associated optimal mapping, in time $\mathcal{O}(n_1^2n_2^2)$. 
\end{theorem}
\begin{proof}
    All entries of the vector $\delta$ for the optimal mapping are critical values of the weak graph distance. The first entry of the vector always corresponds to the weak graph distance, which we compute in the first two steps using the algorithm from \cite{akitayaetal21} in time $\mathcal{O}(n_1n_2\log(n_1+n_2))$ and correctly store as the first value of our distance vector.
	If we remove the corresponding bottleneck using the procedure described, then, by Lemma \ref{Lemma-Korrektheit-Eliminierung}, we obtain modified graphs $G_1$ and $G_2$, whose distance vector $\tilde{\delta}$ corresponds to the vector $\delta$ without its first entry. In particular, $\tilde{\delta_1}$ is the weak graph distance of the current graphs. If we therefore continue to iterate through all possible critical values $\varepsilon$, we find all entries of the distance vector $\delta$ and thus, implicitly, an optimal mapping that realises the lexicographic graph distance. 
	For each of the $\mathcal{O}(n_1n_2)$ critical values, the decision algorithm with a runtime of $\mathcal{O}(n_1n_2)$ is executed at most once.
    We therefore obtain a runtime of $\mathcal{O}(n_1^2n_2^2)$.
\end{proof}

\newpage
\setcounter{theorem}{4}
 \section{Edge-based optimal weak graph distance} 
 We formalise the idea of optimising the mapping by minimising all weak \F distances between the edges and their images:
  \begin{definition}
     Let $s\colon G_1\rightarrow G_2$ be a mapping. Let $e_1,\dots,e_m$ be an ordering of the edges such that  $\delta_1\geq\dots\geq\delta_m$, where $\delta_i=\delta_{wF}(e_i,s(e_i))$. We call $(\delta_1,\dots,\delta_m)$ the \textbf{edge-based distance vector} of $s$. 	
     A mapping $s$ with edge-based distance vector $\delta$ is called \textbf{edge-based optimal} if, for every other mapping $\tilde{s}$ with distance vector $\tilde{\delta}$, we have $\delta\leq_{lex}\tilde{\delta}$.
 \end{definition}
 
  Since the weak graph distance is the smallest possible maximum distance between an edge and its image, the first entry $\varepsilon_1$ of the distance vector of an edge-based optimal mapping corresponds exactly to the weak graph distance. The remaining edges are then mapped in such a way that, once again, the maximum required weak \F distance between edges and their images is minimised. To compute the edge-based distance vector and thus the edge-based optimal mapping between two graphs, we can repeatedly apply the computation algorithm for the weak graph distance and, after each computation, adjust the graph $G_1$ in such a way that the current bottleneck is eliminated and the optimal mapping can be determined for the remaining graph. 
  However, unlike in the algorithm from \cite{akitayaetal21} for the weak graph distance, we do not want to recompute all possible vertex placements at every step. Instead, for each vertex $v\in V_1$, we maintain a list $L_v$ of its vertex placements that are still valid. From this list, the vertex placements valid for the new critical value can then be computed the next time the computation algorithm is run.  
  For convenience, when we apply the modified computation algorithm to compute the weak graph distance with restricted vertex placements, we simply refer to this as computing the weak graph distance as well.  

\begin{figure}
    \begin{center}
        \includegraphics[scale=0.8]{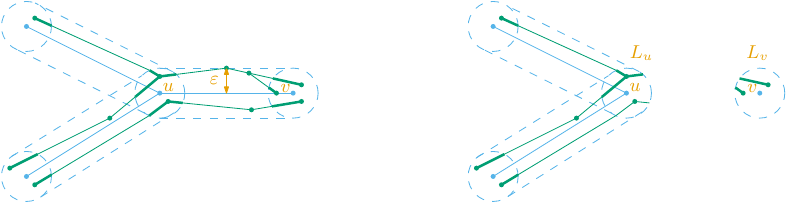}
    \end{center}
    \caption{Parts of two graphs ($G_1$ blue, $G_2$ green), where the weak graph distance is attained on an edge (left) and the vertex $v$ has degree~1. Deleting the edge as bottleneck leaves parts of the graphs (centre), while for $v$ we only store the remaining valid vertex placements. The bold edge segments represent the possible vertex placements. }
    \label{Beispiel_kGA_Kante_Löschen}
\end{figure}

The following observation allows us to remove the bottlenecks from our graphs:  If the weak graph distance $\delta$ is attained between a vertex $v$ on $G_1$ and a point $p$ on $G_2$, then we know the exact point to which vertex $v$ must be optimally mapped. We therefore also know that each of the edges $(u,v)$ incident to $v$ must have a weak \F distance $\delta$ to its optimal image. So we just need to store which placements of $u$ allow an edge placement for these edges; we then do not need to consider the edges any further and can delete them.  
If the distance is attained between a point on an edge $(u,v)$ of $G_1$ (not at a vertex) and a point $p$ of $G_2$, we can store those placements of $u$ and $v$ that allow a valid edge placement through the point $p$, and then delete the edge (see Figure ~\ref{Beispiel_kGA_Kante_Löschen}).

The following pseudocode describes the algorithm for computing an edge-based optimal mapping for the weak graph distance with the corresponding distance vector $\delta$:

  \begin{algorithm}[H]
  	\caption{Edge-based optimal weak graph distance}
    \hspace*{\algorithmicindent} \textbf{Input:} Graphs $G_1$ and $G_2$ 
    \begin{algorithmic}[1]
            \State Compute $\varepsilon=\vec{\delta}_{wG}(G_1,G_2)$.
            \State For each vertex $v\in G_1$, store the vertex placements $L_v$ that are still valid and the connected components to which they belong.
            \State $i \leftarrow 1$, $m \leftarrow |E_1|$
            \State \textbf{Procedure:} Eliminate bottleneck $\varepsilon$
            \While{$i\leq m$}
            \State Compute $\varepsilon=\vec{\delta}_{wG}(G_1,G_2)$ using the list of valid vertex placements.
            \State \textbf{Procedure:} Eliminate bottleneck $\varepsilon$
            \EndWhile
	\end{algorithmic}
    \hspace*{\algorithmicindent} \textbf{Output:} Distance vector $\delta=(\delta_1,...,\delta_m)$, list of valid vertex placements
    \label{Algo:KosG}
   \end{algorithm}

 The computation of the weak graph distance in line $6$ uses the algorithm from~\cite{akitayaetal21}, which is a parametric search using the decision algorithm. We modify the first step of the decision algorithm – that is, the computation of vertex placements –  
 as follows:  
 Instead of recomputing all vertex placements for a critical value $\varepsilon$, we compute the new vertex placements for all vertices still present in the graph $G_1$ from the stored lists $L_v$ by computing the intersection of the stored placements and the new, smaller $\varepsilon$-ball centered on vertex $v$. This is done once for each stored placement, so this step takes $\mathcal{O}(n_1m_2)$ time, as in the original algorithm. 
At the end of the computation of $\varepsilon$, we also overwrite the lists $L_v$ with the placements that are now still valid for each $v\in G_1$ and the connected components to which they belong.
When computing the current weak graph distance, all critical values greater than the last value $\delta_i$ can be discarded immediately, as the distance decreases with each iteration. However, this does not influence the asymptotic running time.

The following pseudocode explains how the current bottleneck is removed from the graph~$G_1$ (following the observation above).
\begin{algorithm}[H]
 	\caption{Procedure: Eliminate bottleneck $\varepsilon$}
 	\begin{algorithmic}[1]
	 		  \If {$\varepsilon$ is the distance from $v\in V_1$ to $G_2$ }
  			\For{$j\leftarrow 0$ to $deg(v)-1$}
            \State $\delta_{i+j}\leftarrow \varepsilon$
            \EndFor
  			\State $i\leftarrow i+deg(v)$
            \State For each vertex $u$ adjacent to $v$, remove from $L_u$ those placements that are not 
            \hspace*{\algorithmicindent} reachable from the placement chosen for $v$.
            \State  Remove $v$ and all edges incident to $v$ from $G_1$.
            \Else {  $\varepsilon$ is attained between a point on $e=(u,v)\in E_1$ and a point $p$ in $G_2$}
            \State $\delta_{i}\leftarrow \varepsilon$
            \State  Delete all vertex placements of $u$ from $L_u$ and of $v$ from $L_v$ that do not belong 
            \hspace*{\algorithmicindent} to the same connected component as $p$.
            \State Delete edge $e$ from $G_1$.
            \State $i++$
            \If{There are no more edges incident to $u$}
            \State Delete vertex $u$.
            \EndIf
             \If{There are no more edges incident to $v$}
             \State Delete vertex $v$.
            \EndIf
  			\EndIf
	 		\end{algorithmic}
            \label{Prozedur: kEB}
\end{algorithm} 

\begin{lemma}\label{Lemma-Korrektheit-Eliminierung-KB}
Let $s\colon G_1 \rightarrow G_2$ with distance vector $\delta = (\delta_1, \ldots, \delta_m)$ be an edge-based optimal mapping from $G_1$ to $G_2$. Then the procedure ‘Eliminate Bottleneck $\delta_1$’ correctly eliminates all $j$ edges from $G_1$ with weak \F distance $\delta_1$ with respect to $s$, such that $\tilde{s}$ with $\tilde{\delta}=(\delta_{j+1},...,\delta_m)$ is edge-based optimal for the residual graph $\tilde{G_1}$ with the stored vertex placements. Furthermore, the first $j$ entries with value $\delta_1$ of the distance vector are stored.
\end{lemma}

\begin{proof}
Each entry in the distance vector represents the weak \F distance of a fixed edge to $G_2$ under an optimal mapping. However, the individual entries are not independent of one another, as the end-vertices of the edges must be mapped consistently. Apart from these end-vertices, however, each edge of $G_1$ is relevant only to the single entry associated with it.
In particular, this means that, in order to eliminate the bottleneck where the value $\delta_1$ is attained, we can remove the corresponding edge $e=(u,v)$ from $G_1$. However, in order not to alter the other entries, we must ensure that, for an optimal mapping of $\tilde{G_1}$, the end-vertices of this edge may only be mapped in such a way that, together with the mapping of the edge $e$, we obtain a consistent optimal mapping for the graph $G_1$ as a whole.

We therefore remove from our lists $L_u$ and $L_v$, for the end-vertices $u$ and~$v$, those vertex placements with which a mapping of the edge $e$ with maximum weak \F distance $\delta_1$ would not be possible. For a more detailed analysis, we distinguish between the two cases where $\delta_1$ is attained inside an edge or at a vertex of $G_1$ (see Figure~\ref{Beispiel_kGA_Kante_Löschen}):

\begin{itemize}
   \item The weak graph distance $\delta_1$ is attained between a vertex on an edge $(u,v)$ of $G_1$ and a vertex $p$ of $G_2$. This means that, in an optimal mapping, the weak graph distance of the edge $(u,v)$ to its image is equal to $\delta_1$. 
    We therefore remove the edge $(u,v)$ from $G_1$, thereby eliminating the distance $\delta_1$ from the distance vector. It remains to be shown that the remaining distance vector is not changed. To ensure that, for the entire mapping, only paths from $G_2$ with the minimum weak \F distance to $(u,v)$ are used, only those vertex placements of $u$ and $v$ may be used in further calculations which, for $\delta_{wG}$ (for the current distance) lie in the same connected component as our point $p$ and thus allow an edge placement for $(u,v)$ via $p$. We therefore remove from the lists $L_u$ and $L_v$ those placements that belong to other connected components. The rest of the graph $G_1$ remains unchanged, so the rest of the distance vector $\tilde{\delta}=(\delta_2,...,\delta_m)$ remains the same. Since we assume that the critical values used to calculate the weak graph distance are different, we have $\delta_1\not = \delta_2$.
    \item The weak graph distance $\delta_1$ is attained at a vertex $u$ in $G_1$. This means that we have found a specific placement $p$ for our vertex $u$, and the weak distance of each of the $j$ edges incident to $u$ to their image $\delta_{wG}$ is equal, so $\delta_1=...=\delta_j$. We therefore remove vertex $u$ along with all $j$ incident edges $(u,v)$ from $G_1$, thereby also eliminating the first $j$ entries of the distance vector $\delta$. 
    Furthermore, for all these vertices $v$ we also remove those placements that lie in a different connected component (in the tube of the edge with the current distance $\varepsilon$) to $p$, in order to guarantee a consistent optimal mapping. The rest of $G_1$ remains unchanged, and adjusting the vertex placements does not change the remaining distance vector. It holds that $\tilde{s}$ with $\tilde{\delta}=(\delta_{j+1},...,\delta_m)$ is edge-based optimal for $\tilde{G_1}$ and $G_2$. Furthermore, $\delta_{j}\not =\delta_{j+1}$.
\end{itemize}
In total, therefore, the $j$ identical first values of the distance vector $\delta$ are correctly determined and stored. Furthermore, $G_1$ and the stored vertex placements are adjusted such that $\tilde{s}$ is edge-based optimal for the remaining graph $\tilde{G_1}$ with the stored vertex placements, given $\tilde{\delta}=(\delta_{j+1},...,\delta_m)$. The claim therefore follows.
\end{proof}
Using this lemma, we can now prove theorem \ref{wichtig}:

\setcounter{theorem}{2}
\begin{theorem}
    Algorithm~\ref{Algo:KosG} computes an edge-based optimal graph mapping with associated distance vector $\delta$ in $\mathcal{O}(n_1^2n_2\cdot \log(n_1+n_2))$ time. 
\end{theorem}
\begin{proof}
The correctness of the algorithm follows from Lemma~\ref{Lemma-Korrektheit-Eliminierung-KB} and the observation that the first entry of a distance vector corresponding to an optimal mapping from $G_1$ to $G_2$ is equal to the weak graph distance between the two graphs.
The runtime of the ‘Eliminate Bottleneck $\varepsilon$’ procedure is dominated by the deletion of those vertex placements that do not belong to the connected component of the current bottleneck (lines 6 and 10). Since the algorithm from \cite{akitayaetal21} stores the placements along with their connected components, and there can be at most $\mathcal{O}(m_2)$ connected components amongst the $\mathcal{O}(n_1)$ neighbouring vertices, the total running time of the bottleneck elimination procedure is $\mathcal{O}(n_1n_2)$.

The running time of the algorithm is dominated by the computation of the current weak graph distance in line~6, which is $\mathcal{O}(n_1n_2 \cdot \log(n_1+n_2))$ according to \cite{akitayaetal21}. Our modification of the vertex placement computation does not increase this running time, since it only restricts the graph from which the vertex placements are computed.
In each iteration of the algorithm, at least one edge of $G_1$ is eliminated. Hence, at most $m_1$ iterations are required, yielding a total running time of $\mathcal{O}(m_1n_1n_2\cdot \log(n_1+n_2))=\mathcal{O}(n_1^2n_2\cdot \log(n_1+n_2))$.
\end{proof}
\end{document}